\documentclass[namedreferences]{solarphysics}
\usepackage[hyperref,optionalrh,solaromanenum]{spr-sola-addons} % For Solar Physics
\usepackage{graphicx}                    % For eps figures, newer & more powerfull
\usepackage{amssymb}                    % useful mathematical symbols
\usepackage{color}                       % For color text: \color command
\usepackage{hyperref}

\newcounter{defcounter}
\newcommand{\rr}[1]{\textrm{#1}}

\ifx \urlurl    \undefined \def
  \urlurl#1{\href{http://#1}{\textsf{#1}}}\fi

\newcommand{\aap}{{\it Astron. Astrophys.}}

\newcommand{\apj}{{\it Astrophys. J.}}

\begin{document}

\begin{article}

\begin{opening}

\title{A Helicity-Based Method to  Infer the CME Magnetic Field Magnitude in Sun and Geospace:
Generalization and Extension to Sun-Like/M-Dwarf Stars and Implications for Exoplanet Habitability}

%%%%%%%%%%%%%%%%%%%%%%%%%%%%%%%%%%%%%%%%%%%%%%%%%%%
%% Authors Names
%
\author{S.~\surname{Patsourakos}$^{1}$ \sep M.K.~\surname{Georgoulis}$^{2}$}
%-------------------------------------------------
% Runningheads
\runningauthor{S. Patsourakos, M.K. Georgoulis}
\runningtitle{Interplanetary CMEs and Exoplanet Habitability}
%-------------------------------------------------
%% Affilations
\institute{ $^{1}$ Section of Astrogeophysics, Department of Physics, University of Ioannina, 45110 Ioannina, Greece, spatsour@cc.uoi.gr\\
            $^{2}$ Research Center for Astronomy and Applied Mathematics, Academy of Athens, Athens, Greece, manolis.georgoulis@academyofathens.gr  \\}

%%%%%%%%%%%%%%%%%%%%%%%%%%%%%%%%%%%%%%%%%%%%%%%%%%%
%% Runningheads
%
%\runningauthor{}
%\runningtitle{}

%%%%%%%%%%%%%%%%%%%%%%%%%%%%%%%%%%%%%%%%%%%%%%%%%%%
%% Affilations
%% id shold be the same with \author addressref value.
%\address[id={}]{}

%%%%%%%%%%%%%%%%%%%%%%%%%%%%%%%%%%%%%%%%%%%%%%%%%%%
\begin{abstract}
In Patsourakos \textit{et al.} (\apj{} \textbf{817}, 14, \citeyear{pats2016})
and Patsourakos and Georgoulis (\aap{} \textbf{595}, A121, \citeyear{patsgeor2016})
we have introduced a method to infer the axial magnetic field
in flux-rope coronal mass ejections (CMEs) in the solar corona and further away in the interplanetary medium. The method,
based on the conservation principle for magnetic helicity, uses as input estimates  the relative magnetic helicity of the solar source region,
along with the radius and length of the corresponding CME flux rope. The method was initially applied to cylindrical force-free flux ropes, with encouraging results. We hereby extend our framework along two distinct lines. First, we generalize our formalism to several
possible flux-rope configurations (linear/nonlinear force-free, non force-free, spheromak, torus), to
investigate the dependence of the resulting CME axial magnetic field on input parameters and the employed flux-rope configuration. Second, we generalize
our framework to both Sun-like and active M-dwarf stars hosting superflares. In a qualitative sense,
we find that Earth may not experience severe, atmosphere-eroding, magnetospheric compression even for eruptive solar superflares
with energies $\approx 10^4$ times higher than those of the largest \textit{Geostationary Operational Environmental Satellite} (GOES)
X-class flares currently observed. In addition, the two recently discovered exoplanets with the highest Earth similarity index, Kepler 438b and Proxima b, seem to lay in the prohibitive zone of atmospheric erosion due to interplanetary CMEs (ICMEs), except in case they possess planetary magnetic fields that are much higher than Earth's.
\end{abstract}

%%%%%%%%%%%%%%%%%%%%%%%%%%%%%%%%%%%%%%%%%%%%%%%%%%%
%% Keywords
%
%\keywords{}

\end{opening}
%-------------------------------------------------

%%%%%%%%%%%%%%%%%%%%%%%%%%%%%%%%%%%%%%%%%%%%%%%%%%%
%% Sections
%
\section{Introduction}

Knowledge of the magnetic field of coronal mass ejections (CMEs), both near the Sun, and further away at 1 AU,
is a key parameter toward understanding their structure, evolution, energetics, and  geoeffectiveness.
For example, the energy stored in  non-potential CME magnetic fields is more than sufficient to counterbalance their mechanical energy \citep[\textit{e.g.}][]{forbes2000,av2000}. In addition, the magnitude of the southward interplanetary (IP) magnetic field associated with CMEs reaching 1 AU is arguably the most important parameter for determining the magnitude of the associated geomagnetic storms  \citep[\textit{e.g.}][]{wu2005}. In stellar obsevations, the recent detections of superflares  with energies of up to $10^4$ times the energy of  ``typical" solar flares,  on Sun-like stars
\citep[\textit{e.g.}][]{maehara2012,shibayama2013}, and their possible association with stellar CMEs, may have significant implications for the physical conditions and the eventual habitability of exoplanets
orbiting super-flaring stars \citep[\textit{e.g.}][]{khoda2007,lammer2007,vidotto2013,armstrong2016,kay2016}.
The stellar CME magnetic field at exoplanet orbit is an important parameter for its habitability since
it enters into  the calculation of the exoplanet magnetosphere size.

Unfortunately, few direct determinations of near-Sun CME magnetic fields currently exist, not to mention
stellar CMEs,
\citep[\textit{e.g.}][]{bastian2001,jensen2008, tun2013, hariharan2016, howard2016, kooi2016}. These determinations occur from non-routine, exceptional observations in the radio domain (\textit{e.g.}
Faraday rotation, moving type IV  bursts).

In parallel,  methods for CME magnetic field inference have emerged  \citep[][]{kunkel2010,savani2015}.
We have recently developed a novel method to infer the near-Sun and 1 AU magnetic field magnitude. For the remainder
of this article, and for brevity, by CME magnetic field we will mean CME magnetic field magnitude.
\citep{pats2016,patsgeor2016}. In a nutshell, our method is based on the principle
of magnetic helicity conservation applied to CMEs. It uses analytical relationships
connecting
the CME magnetic field  with its magnetic helicity
$H_\rr{m}$, and a set of its geometrical parameters (\textit{e.g.} CME
length, radius). Magnetic helicity is a signed quantity, depending
on its handedness. For the remainder of this article, and for brevity, when referring to magnetic helicity we will
refer to its magnitude.

CME magnetic helicity and geometrical parameters
can be routinely deduced from
photospheric and coronal observations, respectively.  Invoking the magnetic helicity conservation property
\citep[\textit{e.g.}][]{berger1984}, therefore, allows
to infer the CME near-Sun
magnetic field. Finally, radial power-law extrapolation of the {\bf inferred}
near-Sun CME magnetic field allows to calculate its value at 1 AU.
We  applied our method to an observed case-study in \citet{pats2016},
corresponding to a super-fast CME which gave rise to one of the most intense geomagnetic
storms of Solar Cycle 24. In  \citet{patsgeor2016} we performed a parametric
study of our method, based on the observed distributions of its input parameters
(CME magnetic helicity and geometrical parameters). We used
one of the most common  flux-rope CME models (Lundquist 1950).

In the present study we extend our initial work in two meaningful ways: first, we generalize
our parametric study on an array of proposed theoretical CME models, thereby further constraining the
radial field evolution of the modeled CMEs in the interplanetary medium
(Sections 2 and 3). Second, we extend our framework to non-solar
cases, particularly to stars hosting superflares,
and determine the magnetic fields of possible
stellar CMEs associated with these extreme events.
This allows us to infer some rough limits on magnetosphere size of  exoplanets orbiting such stars and to cast some preliminary implications for their habitability (Section 4). We conclude with a summary and a discussion of our results, as well as possible avenues for future research (Section 5).

\section{The Helicity-Based Method to Infer the Near-Sun  and 1 AU CME Magnetic FIeld}
We briefly describe here our framework for inferring the near-Sun  and 1 AU  magnetic field of CMEs, in Sections 2.1 and 2.2, respectively,
and its parametrization, in Section 2.3. More details can be found in  \citet{pats2016} and \citet{patsgeor2016}.

\subsection{Determination of CME Near-Sun Magnetic Field}
First, we  infer  the near-Sun CME magnetic field as follows:
\begin{enumerate}
\item Determine the magnetic helicity $H_\rr{m}$ of the CME solar source region using
various methods based on theory and photospheric observations  \citep[][among others]{pariat2006,regn2006,geor2012,val_etal12,morait2014}.
\item Attribute the source-region $H_\rr{m}$ to the analyzed CME.
\item Determine a set of CME geometrical parameters (\textit{e.g.} radius $R$, length $L$)
from forward-modeling geometrical fits of multi-view coronal observations of the analyzed CME  (Section A of Appendix).
Such observations are achieved by coronagraphs typically covering the outer corona.
\item Plug the magnetic  and geometrical parameters of the CME deduced in the previous two steps
into theoretical formulations, such as the  models described in Sections B-G of  Appendix, and deduce the
corresponding near-Sun magnetic field magnetic field $B_{*}$ at heliocentric distance $r_{*}$.

\end{enumerate}

\subsection{Extrapolation of Near-Sun CME Magnetic Field to 1 AU}
Next, we extrapolate $B_{*}$ from $r_{*}$, where coronagraphic observations
of the CME geometrical properties are taken, outward in the IP space and eventually to 1 AU. We assume that its radial evolution is described by a power-law in the heliocentric radial distance $r$:
\begin{equation}
B_{0}(r)=B_{*} {(r/r_{*})}^{{\alpha}_{B}}.
\label{eq:scaleb}
\end{equation}
In Equation (\ref{eq:scaleb}) we assume that the power-law index ${\alpha}_{B}$ varies in the
range [-2.7, -1.0]. This
results from various theoretical
and observational studies \citep[\textit{e.g.}][]{patzold1987,kumar96,bothm98,vrasn04,liu05,forsh06,leitn07,dem2009,moestl2012,poomv2012,mancuso2013,good2015,
winslow2015}.
Notice here that most of these studies do not fully cover the  range  we are considering here,
but typically subsets thereof, either near-Sun or inner heliospheric. We considered 18 equidistant ${\alpha}_{B}$-values and a step of 0.1.
In \citet{pats2016} we applied our method to an observed super-fast CME which gave rise to one of the most
intense geomagnetic storms of Solar  Cycle 24. We found that a rather steep CME magnetic field evolution
of the inferred near-Sun CME magnetic magnetic field in the IP space, with an index  ${\alpha}_{B} \simeq -2$,
was consistent with the range of the associated interplanerary CME (ICME) magnetic field values as observed in-situ at $\approx$1 AU
(\textit{i.e.} at the L1 Lagrangian libration point).

\subsection{Parametric Study of the Helicity-Based Method}
In \citet{patsgeor2016} we performed a parametric study of our method. We
carried out Monte-Carlo simulations of ${10}^{4}$ synthetic CMEs and used the frequently-used
Lundquist model (Section B of Appendix).
Our simulations  randomly sampled ${10}^{4}$  values of active region (AR) $H_\rr{m}$, CME aspect ratio $k$ and
angular half-width $w$ from the observed distributions of 42 ARs   \citep[][]{tzio2012} and 65  CMEs \citep[][]{thern2009,bosman2012}, respectively.
Using the randomly selected CME aspect ratio and angular half-width in Equations  (\ref{eq:cmer}) and
(\ref{eq:cmelength}), the CME radius and length, respectively, were deduced, and finally, the
CME magnetic field was determined via Equations (\ref{eq:hm}) and (\ref{eq:alpha}).
As a result,  $10^4$ $B_{*}$ values at $r_{*}=10\,\mathrm{R_{\odot}}$ were calculated.
These near-Sun magnetic fields were then extrapolated to 1 AU,
therefore leading to 180,000 1 AU CME magnetic field values, resulting from the matrix of ${10}^{4}$ $B_{*}$ and 18 ${\alpha}_{B}$ values discussed in Sections 2.1 and 2.2.
We found  that an  index ${\alpha}_{B}$ = -1.6 $\pm 0.2$ led to a ballpark, statistical agreement
between  the model-predicted ICME magnetic field distributions
and actual ICME observations at 1 AU.

\section{Parametric Study of the Helicity-Based Method for Different CME Models}
In this Section we extend and generalize the parametric study of  \citet{patsgeor2016} to six
different CME models, including the Lundquist model used in that study. These models,
described in Sections B-G of  Appendix, make different assumptions about the
CME shape (cylindrical segment, toroidal, spheromac), the nature of their currents (linear/nonlinear force-free, non force-free) and the distribution of twist (uniform, non-uniform), thereby allowing maximum flexibility in our method's application.

As in   \citet{patsgeor2016}, we perform Monte-Carlo simulations picking up ${10}^{4}$ random deviates from
the observed distributions of the input magnetic ($H_\rr{m}$) and geometrical ($k$ and $w$) parameters of our method
and determine ${10}^{4} B_{*}$  values for each considered model using the corresponding
equations per model (Sections B-G of  Appendix). Depending on the specific model assumptions, extra geometrical parameters
are used (\textit{e.g.} CME minor and major radius in  Equation (9); number of turns in Equations (11) and (13)). For each considered model we calculate
the probability density function (PDF) of: (1) the extrapolated to 1 AU CME magnetic fields, ${w{B}_\rr{1 AU}}$, for the    ${10}^{4}$ $B_{*}$ values and for each  of the 18 considered  ${\alpha}_{B}$
values, $\rr{{PDF}}_{\rr{mod}}$; and (2) the magnetic fields, $B_\rr{MC}$,  for 162 magnetic clouds (MCs) observed at 1 AU as
resulting from their linear force-free fits \citep[][]{lynch2003,lepping2006}, and $\rr{{PDF}}_\rr{obs}$.

\begin{figure}
\centerline{\includegraphics[width=0.9\textwidth,clip=]{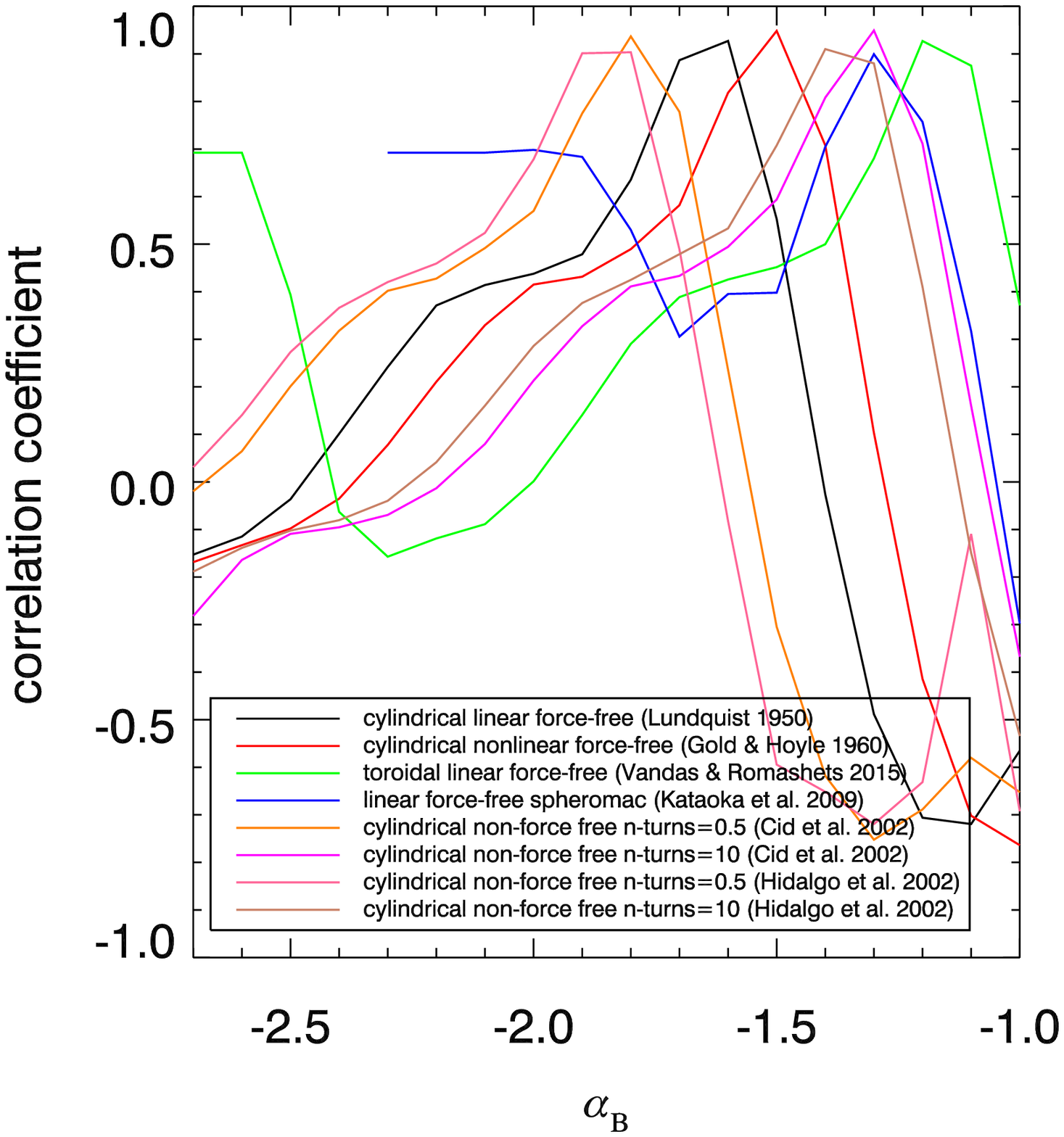}}
\caption{Correlation coefficient of the PDFs for the predicted $B_\rr{1AU}$  and observed $B_\rr{MC}$ values at L1 as
a function of ${\alpha}_{B}$ for the various employed models.
} \label{fig:fig1}
\end{figure}

\begin{figure}
\centerline{\includegraphics[width=0.9\textwidth,clip=]{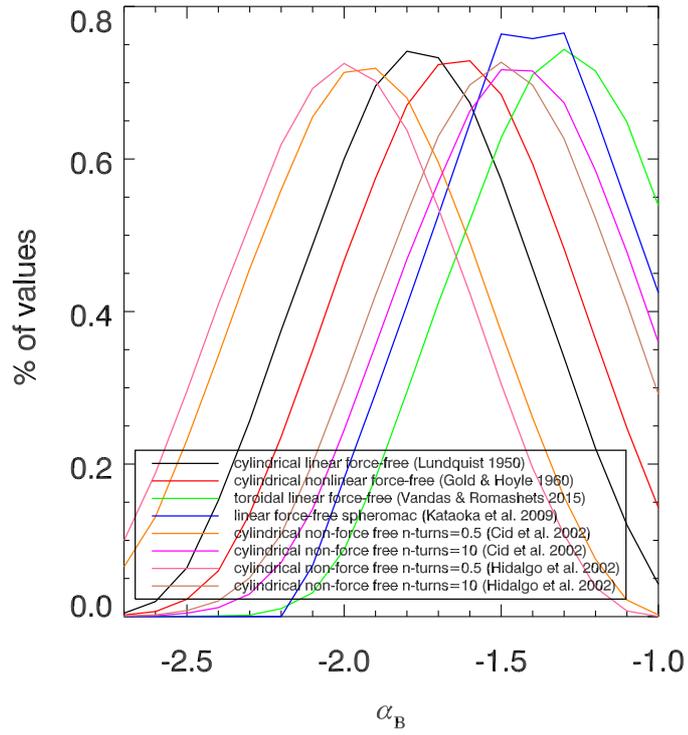}}
\caption{Fraction of ${B_\rr{1 AU}}$ values falling within the
observed $B_\rr{MC}$ range of 4\,--\,45 nT as a function of ${\alpha}_{B}$ for the various employed models.} \label{fig:fig2}
\end{figure}

We compare the results of our Monte-Carlo simulations for the six different employed models with  MC observations in Figures \ref{fig:fig1} and
\ref{fig:fig2}.  In   Figure \ref{fig:fig1} we display the correlation coefficient of
$\rr{{PDF}}_\rr{mod}$ and  $\rr{{PDF}}_\rr{obs}$
as a function of ${\alpha}_\rr{B}$, whereas  in Figure \ref{fig:fig2}  we diplay
the fraction $\rr{{frac}}_\rr{MC}$ of ${B_\rr{1 AU}}$ values falling within the
observed $B_\rr{MC}$ range, namely [4,45] nT,  as a function of ${\alpha}_{B}$.

Several remarks can be now made:

\begin{enumerate}
\item Both $\rm{{PDF}}_\rr{mod}$ versus $\rm{{PDF}}_\rr{obs}$ correlation coefficients and
$\rr{{frac}}_\rr{MC}$
exhibit well-defined peaks, and cover similar ${\alpha}_{B}$-ranges of $\approx$ [-1.85, -1.2]
and [-2.0,-1.3], respectively, for all considered models. For
the cases of the toroidal and spheromak linear force-free models there are
secondary peaks for ${\alpha}_{B} < -2$ (Figure \ref{fig:fig1}). However
these peaks are not supported by high $\rr{{frac}}_\rr{MC}$-values.
\item Similar (relatively high) peak values can be found in both Figures
\ref{fig:fig1} ($\approx$ 0.9) and \ref{fig:fig2} ($\approx$ 0.7) for all considered models. The full width at half maximum
(FWHM) of the curves of Figures
\ref{fig:fig1} and \ref{fig:fig2} is $\approx$ [0.3,0.5] and [0.7,0.8], respectively, in the ${\alpha}_\rr{B}$-range.
\item The linear and
nonlinear force-free, cylindrical, flux-rope models, black and red lines, respectively, in both Figures \ref{fig:fig1}
and \ref{fig:fig2}, place  the
median of the considered models.
\end{enumerate}

Figure \ref{fig:fig3} displays the model-averaged $\rr{{PDF}}_\rr{mod}$ versus $\rr{{PDF}}_\rr{obs}$ correlation coefficient
and $\rr{{frac}}_\rr{MC}$  as a function of ${\alpha}_{B}$.  These
plots encapsulate the overall performance
of all considered models. The corresponding curves  peak at ${\alpha}_{B} \approx -1.6$ and
${\alpha}_{B} \approx -1.8$ for the  $\rr{{PDF}}_\rr{mod}$ versus $\rr{{PDF}}_\rr{obs}$ correlation coefficient
and  $\rr{{frac}}_\rr{MC}$,  respectively. Both peaks correspond to relatively high values of $\approx$0.6.

\section{Extension to Stellar CMEs}
Existing stellar observations lack the spatial resolution to directly image
possible stellar CMEs, not to mention their source regions.
Therefore, we have to rely on indirect observational inferences for stellar CMEs
\citep[\textit{e.g.}][]{houdebine1990,leitzinger2011,osten2013,leitzinger2014}. Solar
observations suggest that with increasing flare magnitude, the probability of
an associated CME is also increasing, with GOES-class flares $\gtrsim$X3 essentially associated one-to-one to fast or superfast CMEs \citep[\textit{e.g.}][]{andrews2003,yashiro2005,nindos2015,harra2016}.
Therefore, it is possible that stars with superflares could have enhanced rates of superfast CMEs associated to them. Stars hosting superflares are generally characterized by enhanced
magnetic activity, as inferred from their larger chromospheric emissions
\citep[\textit{e.g.}][]{karoff2016}. A common practice in studies of stellar flares and CMEs is to extrapolate empirical relationships derived from solar studies to the stellar context \citep[\textit{e.g.}][]{aarnio2012, drake2013, osten2015}.

Therefore, in order to determine the $H_\rr{m}$ associated with stellar CMEs, we rely on a recently discovered empirical relationship connecting
the free coronal magnetic energy $E_\rr{f}$
and $H_\rr{m}$ of  solar active regions. \citet{tzio2012} inferred the following best fit
for 42 solar ARs and 162 snapshots thereof:
\begin{equation}
\rr{log}|H_\rr{m}|=53.4 - 0.0524{(\rr{log}(E_\rr{f}))^{0.653}}\rr{exp}(\frac{97.45}{\rr{log}(E_\rr{f})}).
\label{eq:hmefree}
\end{equation}
Equation (\ref{eq:hmefree}) was applied to an $\rm{E_f}$ interval of ${10}^{30}-{10}^{33} \mathrm{erg}$.
We consider stellar eruptive flares (\textit{i.e.} assosiated
with stellar CMEs) with energies in the range ${10}^{32}-10^{36} \mathrm{erg}$, and cover
this interval with a  logarithmic step of 0.1 leading to 41(=$N_{E_\rr{f}}$) bins in $E_\rr{f}$.
The employed interval includes both  solar flares and  stellar superflares \citep[\textit{e.g.}][]{shibayama2013};
stellar superflares have a most probable energy of few times ${10}^{34}$ erg.
Notice that typically the quoted superflare energies are bolometric, \textit{i.e.} correspond
to the wavelength-integrated radiated energy \citep[\textit{e.g.}][]{maehara2012,shibayama2013}.
For each considered flare energy we assume an equal free magnetic energy.
This assumption imposes a lower limit on the free magnetic energy of the host active region,
since we know that the largest flare in an active region releases a relatively small percentage (typically
10\,--\,20\,\%)  of the free magnetic energy stored in the region \citep[\textit{e.g.}][]{lynch2008,sun2012,tzio2012, tzio2013}.

Having determined the magnetic helicities that could be associated with stellar CMEs, we then
adopt typical values for $k$ (=0.3) and  $w$ (=20 degrees) as resulting from
solar studies of CMEs \citep[\textit{e.g.}][]{thern2009,bosman2012} and determine their near-star
stellar CME magnetic
field values at 10 $\mathrm{R_{\odot}}$
by applying the Lundquist model (Section B of Appendix). Next,  adopting
the radial power-law CME magnetic field evolution of Equation (\ref{eq:scaleb}), we
extrapolate  the derived near-star stellar CME magnetic fields in the range [0.05,1.5] AU, which spans
the habitable zone (HZ) of exoplanet candidates  \citep[][]{kane2016}. HZ is typically defined
as the distance from a mother star at which water could exist in a liquid form
\citep[\textit{e.g.}][]{kasting1993,selsis2007,khoda2007}. This obviously corresponds
to a minimum requirement for life-sustaining conditions in exoplanets.
The HZ migrates closer to the mother star when moving from Sun-like
stars to M-dwarfs \citep[\textit{e.g.} see Figure 3 in][]{selsis2007}.

We cover the above-mentioned radial distance  interval
with a radial step of  0.003 AU, leading to 480($=N_{r}$) bins in radial distance.
We assume that this interval is populated with exoplanets having a magnetosphere.
For the extrapolation we employ
an index ${\alpha}_{B}=-1.6$ since it corresponds
to the  ``best-fit" for the Lundquist model
for the solar-terestrial case \citep[][]{patsgeor2016}.
We therefore calculate a two-dimensional, $41 \times 480$ $N_\rr{{r}}$ versus $N_{E_\rr{{f}}}$ grid
of extrapolated stellar CME magnetic fields, at the locations of
hypothetical exoplanets and for the employed stellar flare energies.
We finally determine the magnetopause radius of  hypothetical
exoplanets at the considered distance from
a pressure-balance equation, balancing the stellar CME magnetic pressure at the exoplanet's
vicinity with the magnetic pressure of the intrinsic, assumed
dipolar, planetary magnetic field \citep[\textit{e.g.}][]{chapman1930}:
 \begin{equation}
\frac{B_\rr{CME}^{2}}{8\pi}=\frac{B_\rr{eq}^{2}}{8\pi}{(\frac{1}{r_\rr{mpause}})}^{6}.
\label{eq:mpause}
\end{equation}
In Equation  (\ref{eq:mpause}), $B_\rr{CME}$ is the extrapolated stellar CME magnetic
field at the vicinity of the exoplanet, $B_\rr{eq}$ is the planetary equatorial
magnetic field, and $r_\rr{mpause}$ is the
magnetopause radius, assumed a dimensionless number in planetary radius ($R_\rr{p}$) units. In writing Equation (\ref{eq:mpause})
we assume a spherical magnetosphere.
Given CMEs are magnetically-dominated
structures, neglecting their thermal and ram pressure
in Equation (\ref{eq:mpause}) can be justified.

In Figure \ref{fig:fig4} we depict  a color representation of
the exoplanet magnetopause radius as a function of
exoplanet distance and stellar flare magnitude.
Given that current knowledge of magnetic fields in exoplanets is incomplete,
we first assume that  $B_\rr{eq}$ is equal to the (current-day) terrestrial equatorial
magnetic field ($B_\rr{E}$=0.333 G). Values of $r_\rr{mpause}$ smaller
than 2 $R_\rr{p}$ are saturated with black.
This threshold corresponds to the minimum magnetosphere size
that could prevent atmospheric erosion from stellar CME impacts  \citep{khoda2007,lammer2007,scalo2007} and may be viewed as an additional,
necessary condition for habitability.  The maximum $r_\rr{mpause}$ is $\approx 10 R_\rr{p}$,
close to the unperturbed value for the terrestrial magnetosphere.
From Figure \ref{fig:fig4} we notice that for exoplanets at distances $ \lesssim $  0.1 AU, practically
all considered flare energies lead to magnetospheric compression below the 2 $R_\rr{p}$  ``habitability
threshold" discussed above. Exoplanets at distances  $\gtrsim$  0.4 AU have $r_\rr{mpause} \gtrsim 2 R_\rr{p}$
for all considered flare energies.

Moreover, exoplanets that are relatively close to their mother star could be subject to
tidal locking, \textit{i.e.} synchronization between their rotational and orbital periods,
and generally a decrease in their  rotation rate
\citep[\textit{e.g.}][]{glad1996}. Slowing-down of an exoplanet
leads to a decrease of its magnetic field \citep[\textit{e.g.}][]{griem2004,griem2005}.
In Figures \ref{fig:fig5} and \ref{fig:fig6} we calculated
$r_\rr{mpause}$ for two more cases, $B_\rr{eq}=0.25 B_\rr{E}$ and $B_\rr{eq}=0.061 B_\rr{E}$, respectively.
These values correspond to calculations of planetary magnetic moments of
Earth-like exoplanets in the close vicinity  of M-type dwarf stars subject
to tidal-locking \citep[][]{khoda2007}.
The maximum $r_\rr{mpause}$ is $\approx 6 R_\rr{p}$ and  $\approx 4 R_\rr{p}$, for Figure \ref{fig:fig5}
and \ref{fig:fig6}, respectively.
Decrease of   the exoplanet magnetic field leads to further compression
of its magnetopause radius for a given stellar flare energy, with exoplanets at distances
$ \lesssim $ 0.2 and $  \lesssim $ 0.4 AU, for 0.25 $B_\rr{E}$ and 0.061 $B_\rr{E}$, respectively, with
$r_\rr{mpause} < 2 R_\rr{p}$.  In the case corresponding to 0.25 $B_\rr{E}$ (Figure \ref{fig:fig5})
exoplanets at distances  $\gtrsim$ 1 AU have $r_\rr{mpause} > 2 R_{p}$
for all considered flare energies. Moreover, in the case corresponding to 0.061 $B_\rr{E}$ (Figure \ref{fig:fig6}),
only stellar flares with  energies  up  ten times the lower limit ($={10}^{33}$ erg) of superflares
could result in  $r_\rr{mpause} > 2 R_\rr{p}$ for exoplanets
at distances $\gtrsim $ 0.7 AU.
Exoplanets at 1 AU experience magnetospheric compression below 2 $R_\rr{p}$ only
in the case of the smallest considered exoplanet magnetic field resulting from tidal locking (Figure
\ref{fig:fig6}).

Figures \ref{fig:fig4}\,--\,\ref{fig:fig6} include two case-studies of recently detected
exoplanets orbiting stars hosting superflares. These are Kepler 438b
\citep[][]{torres2015,armstrong2016} and Proxima  b \citep{anglada2016,davenport2016}, orbiting M-type
dwarfs Kepler 438 and Proxima Centauri, respectively.
Kepler 438b (Proxima b) have orbits with semi-major axis   of 0.16 (0.04) AU and their
mother stars exhibit superflares with energies 1.4$\times{10}^{33}$ (${10}^{33}$) erg.
Kepler 438b and Proxima b have the
highest Earth Similarity Index (ESI), 0.88 and 0.87, respectively, amongst
the detected exoplanets to date\footnote{See
\urlurl{phl.upr.edu/projects/habitable-exoplanets-catalog/data}.
}.
The ESI of a given exoplanet is a metric of its degree
of resemblance with Earth in terms of mass, radius \textit{etc}. An ESI
equal to 1 suggests a perfect match of the exoplanet with Earth.
From Figures \ref{fig:fig4}\,--\,\ref{fig:fig6} we gather that for all considered
stellar flare energies and planetary magnetic fields, the hypothesized stellar CMEs
result in severe magnetospheric compression, below $2 R_\rr{p}$ for both
Kepler 438b and Proxima b. This appears to be a strong constraint against habitability of these exoplanets.

We finally investigate the effect of CMEs
associated with potential solar superflares on the terrestrial magnetosphere.
Based on our current understanding
of solar dynamo and magnetic fields,
geological and historical records, and super-flares observed in Sun-like stars
one may not exclude the  rare (\textit{i.e.} over
timescales of several centuries or more) occurrence of relatively small superflares
(energies of up to $\approx {10}^{34}$)
on the Sun  \citep[\textit{e.g.}][]{shibata2013,uso2013,nogami2014,toriumi2017}, although such possibility
has been contested \citep[\textit{e.g.}][]{scri2012,aulanier2013,cliver2014}.
Inspection of Figure \ref{fig:fig4}, more specifically  by considering
a vertical cut at a distance equal to 1 AU, shows that solar superflares
with energies exceeding ${10}^{33}$ and ${10}^{34}$ $\mathrm{erg}$ could lead
to significant compression of the terrestrial magnetosphere at $\approx$ 6 and 5 Earth radii,
respectively, in
the case $\alpha _B=-1.6$ (Figure \ref{fig:fig7}). For a steeper index $\alpha _B=-1.9$, the magnetopause distance stays above 6 Earth radii for a superflare with energy ${10}^{34}$ $\mathrm{erg}$  (Figure \ref{fig:fig8}). Importantly, in no case
does the magnetopause distance become smaller than 2 Earth radii, even for a  ``worst-case" superflare with energy  ${10}^{36}$ erg.

\section{Discussion and Conclusions}

\subsection{Summary of our Findings and Outlook}
In this work, we (a) perform a parametric study of inferring the near-Sun
and 1 AU magnetic field of CMEs using an array of analytical CME models, and
(b) apply our method to stars hosting superflares. A summary of findings is as follows:
\begin{enumerate}
\item All considered CME models lead to predicted CME magnetic fields
at 1 AU which largely recover the distribution and fall within the range of
magnetic cloud observations at 1 AU, as shown in Figures \ref{fig:fig1} and \ref{fig:fig2}, respectively.
\item The best agreement between the predicted
and observed CME  magnetic fields at 1 AU is achieved in a
different $\alpha_{B}$-range for each model. For all considered models
this range corresponds to $\alpha_B \in [-2.0, -1.2]$. No model appears to significantly outperform the others in terms of agreement between extrapolated CME fields and observations at L1.
\item Stellar CMEs associated with flares and superflares with energies ${10}^{32 }\,-\,{10}^{36}$ erg
could compress the magnetosphere of exoplanets with terrestrial magnetic field orbiting at $<0.1$ AU
below the magnetopause-distance threshold of 2 planetary radii (Figure \ref{fig:fig4}). This threshold signals a CME-induced atmospheric erosion. This zone of severe compression  shifts to $<0.3$ AU for flare energies of ${10}^{34}$ erg.
\item  A CME associated with a solar superflare of ${10}^{34}$ erg
may compress our magnetosphere to a magnetopause distance of $\approx$ 5 Earth radii (Figure \ref{fig:fig7}), about half of its unperturbed value. Even an extreme superflare of ${10}^{36}$ erg cannot push the magnetopause distance to a value lower than 2 Earth radii.
\item For exoplanets with weaker magnetic fields than Earth, particularly tidally locked ones, a superflare can cause severe magnetospheric compression below the 2 planetary radii limit at asterocentric distances $<$ 0.3 and 0.4 AU (Figures \ref{fig:fig5} and \ref{fig:fig6}, respectively).
\item Severe compressions of potential magnetospheres below 2 planetary radii are obtained
for exoplanets Kepler 438b and Proxima b (Figures \ref{fig:fig4}\,--\,\ref{fig:fig6}). The close distance of these exoplanets to mother stars Kepler 438 and Proxima Centauri, respectively, is the main reason for this rather strict constraint on these exoplanet's habitability.
\end{enumerate}

Our major conclusion is that all employed CME models, in spite of differences in key properties (distribution of electric currents, geometry
\textit{etc}.) perform similarly at 1 AU, if applied following the corresponding  ``best-fit"
${\alpha}_{B}$ ranges. This suggests that, for statistical space-weather forecasting purposes,
any of these models can be applied. For a more detailed treatment between models,
aiming toward an improved understanding of the CME-ICME transition and physical state, one needs detailed model
comparisons on a case-by-case basis
at 1 AU \citep[\textit{e.g.}][]{riley2004,alhaddad2013}.
Inner heliospheric and coronal in-situ and imaging
observations are also in high demand, hence anticipation is mounting for the upcoming \textit{Solar Orbiter} and \textit{Solar Probe Plus} missions.

A core assumption of the solar-IP part of our analysis
is that the entire  source-region helicity
is attributed to the corresponding CMEs. This
overestimates the helicity shed by CMEs, since one can rather
reasonably expect that only a fraction of the available
$H_{\rr{m}}$ is expelled in single CME events. If the entire helicity content of the source active region was shed in a single event, that would mean zero remaining free energy, which would be untenable with the essentially unchanged, line-tied photosphere. Further, there may not be enough time
to build-up from zero the helicity expelled in homologous eruptions. CMEs shedding only a fraction of
the source-region $H_{\rr{m}}$ can also be reached by
statistical considerations: for instance,
the eruptive  AR $H_{\rr{m}}$ distributions have
most probable values $>{10}^{43} \mathrm{{Mx}^{2}}$, \citep[\textit{e.g.}][]{nindos2004,tzio2013},
while the MC ones at 1 AU are about an order of magnitute smaller  \citep[\textit{e.g.}][]{lynch2005}.
In addition, analysis of the cumulative $H_{\rr{m}}$ shed by CMEs over long intervals covering
large fractions of (or entire) solar cycles, combined with CME occurrence rates, AR characteristics and
MC $H_\rr{{m}}$ values at 1 AU, leads to the conclusion that on average a CME expells $\approx 10-20\,\%$
of the AR $H_\rr{m}$ \citep[][]{devore2000,geor2009,demou2016}.  Next, a handful of case-studies monitored
eruption-associated $H_\rr{{m}}$ changes in the lower solar atmosphere and/or compared the source region
$H_\rr{{m}}$ with the $H_\rr{{m}}$ of the associated MC at 1 AU, and found that the observed CMEs
expel $\approx 10-70\,\%$ of the available $H_\rr{{m}}$
\citep[\textit{e.g.}][]{dem2002,green2002,nindos2003,luoni2005,mandrini2005,kazachenko2009,nakwa2011,tzio2013}.
In practice, however, most of these studies showed that eruption-related $H_\rr{{m}}$ changes
amount to only $\approx 10-40\,\%$ of the available $H_\rr{{m}}$, notwithstanding
the significant uncertainties due to the various methods and
$H_\rr{{m}}$ determinations. Finally, magnetohydrodynamical (MHD) models of CMEs based on largely different initiation
scenarios show  that $\approx 10-40\,\%$ of the available
$H_\rr{{m}}$ leaves the corresponding computational boxes with the simulated CMEs
\citep[\textit{e.g.}][]{macneice2004,gibson2008,kliem2011,morait2014}.

\begin{figure}
\centerline{\includegraphics[width=0.9\textwidth,clip=]{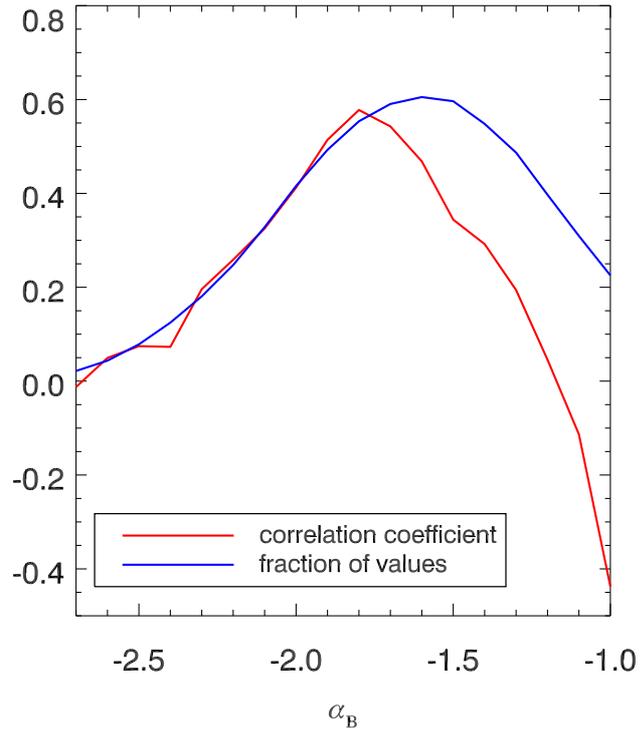}}
\caption{Model-averaged correlation coefficient of the PDF for the predicted $B_\rr{1AU}$ and
observed $B_\rr{MC}$ values at L1 as a function of ${\alpha}_{B}$ (red curve). Also shown is the respective
fraction of ${B_\rr{1 AU}}$ values (blue curve) falling within the
observed $B_\rr{MC}$ range. Both curves correspond to the respective averages of the various employed models as displayed
in Figures \ref{fig:fig1} and \ref{fig:fig2}.} \label{fig:fig3}
\end{figure}

\begin{figure}
\centerline{\includegraphics[width=0.9\textwidth,clip=]{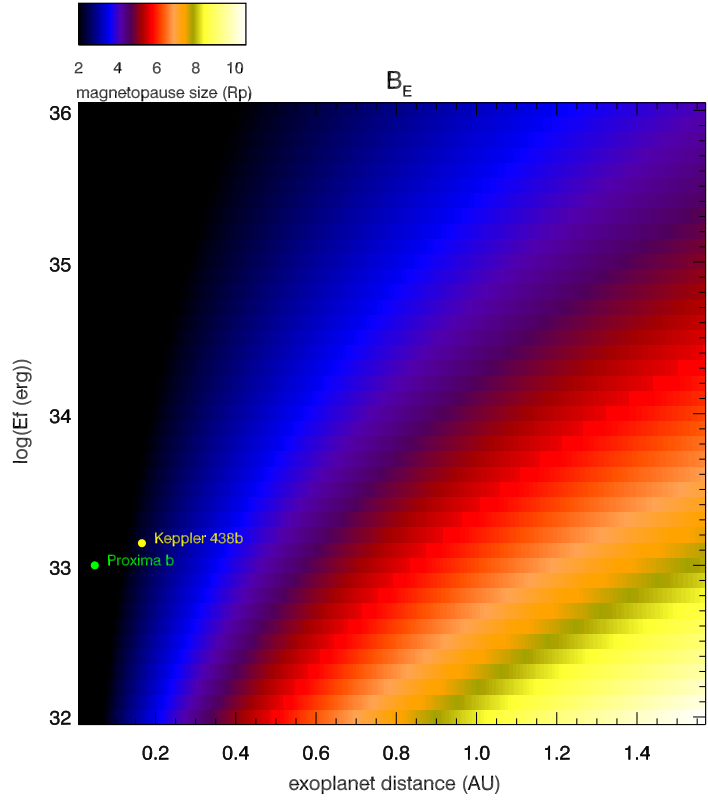}}
\caption{Exoplanet magnetopause radius (in units of planet radius) required to
counterbalance the magnetic pressure of a CME arriving at its vicinity as a function of the
exoplanet distance to, and the associated flare energy of, the planet's mother star.
Magnetopause sizes smaller than 2, representing
a lower limit of magnetospheric size to avoid atmospheric erosion, are saturated with black.
The index ${\alpha}_{B}$  is assumed equal to be -1.6 and the Lundquist model {\bf is} used.
The yellow and green circles correspond to two case-studies
of exoplanets Kepler 438b  and Proxima b, orbiting stars exhibiting superflares.
It is assumed that the exoplanets have an equatorial magnetic field
equal to the terrestrial one, $B_E$ $\approx 0.333$ G.} \label{fig:fig4}
\end{figure}

\begin{figure}
\centerline{\includegraphics[width=0.9\textwidth,clip=]{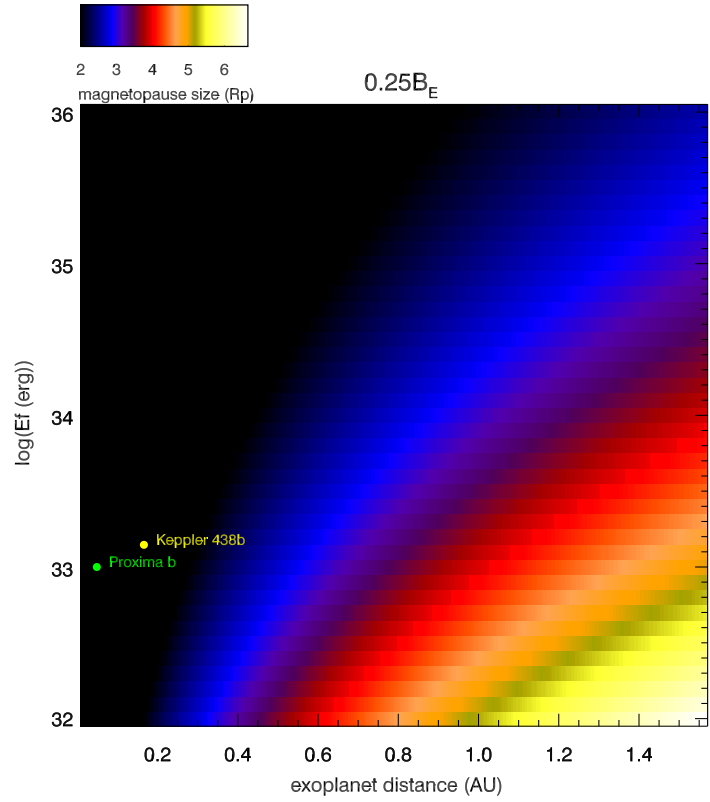}}
\caption{Same as Figure \ref{fig:fig4}, but for an exoplanet equatorial magnetic field of $\approx
0.083$ G, which is  $\approx$ 25\% of the terrestrial value.} \label{fig:fig5}
\end{figure}

\begin{figure}
\centerline{\includegraphics[width=0.9\textwidth,clip=]{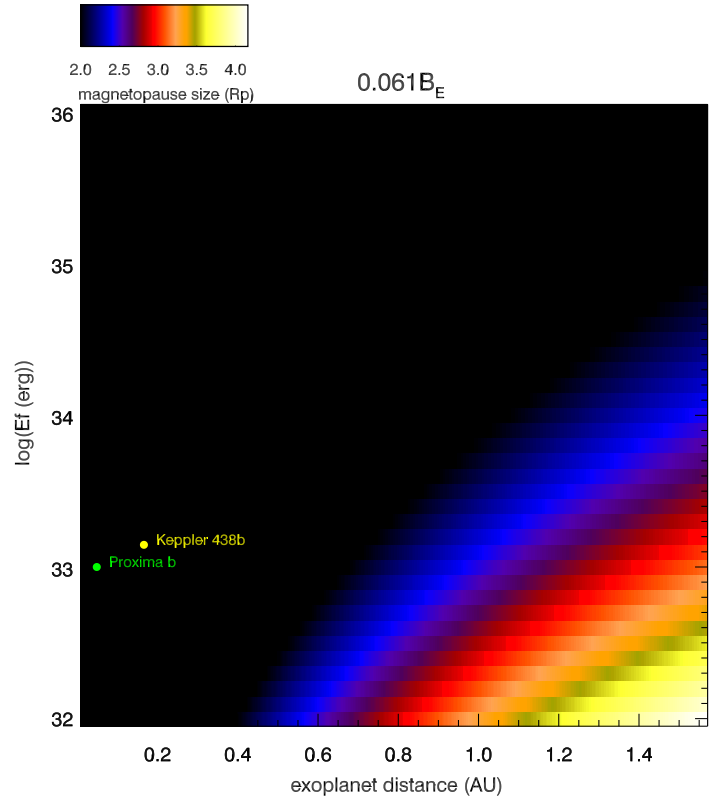}}
\caption{Same as Figure \ref{fig:fig4}, but for an exoplanet equatorial magnetic field of $\approx
0.020$ G, $\approx$ 6.1\% of the terrestrial value.}  \label{fig:fig6}
\end{figure}

\begin{figure}
\centerline{\includegraphics[width=0.9\textwidth,clip=]{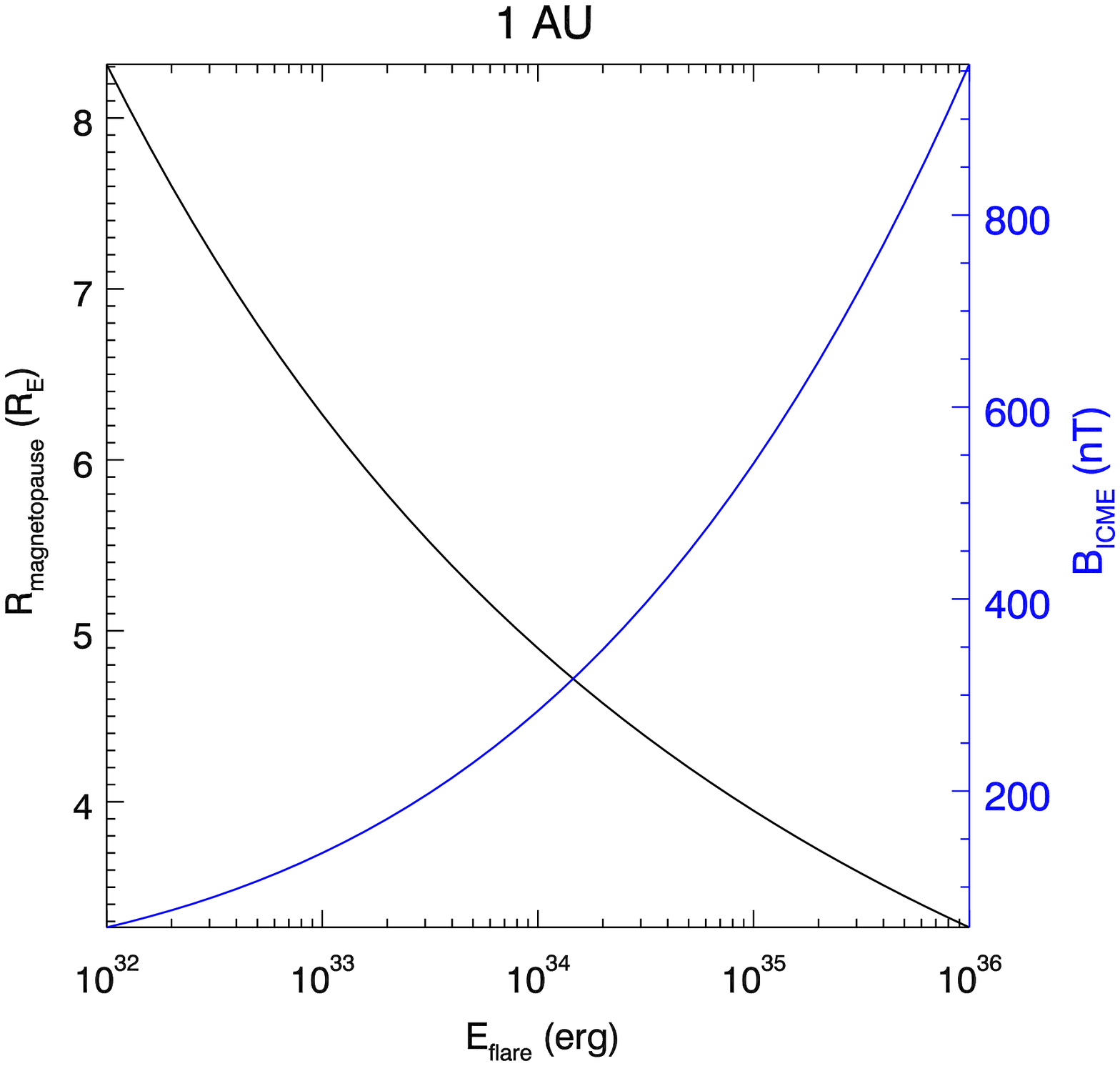}}
\caption{Terrestrial magnetopause radius size (black) and ICME
magnetic field magnitude at 1 AU (blue) as a function of flare
energy. The index
${\alpha}_{B}$ is equal to -1.6 and the Lundquist model is used.} \label{fig:fig7}
\end{figure}

\begin{figure}
\centerline{\includegraphics[width=0.9\textwidth,clip=]{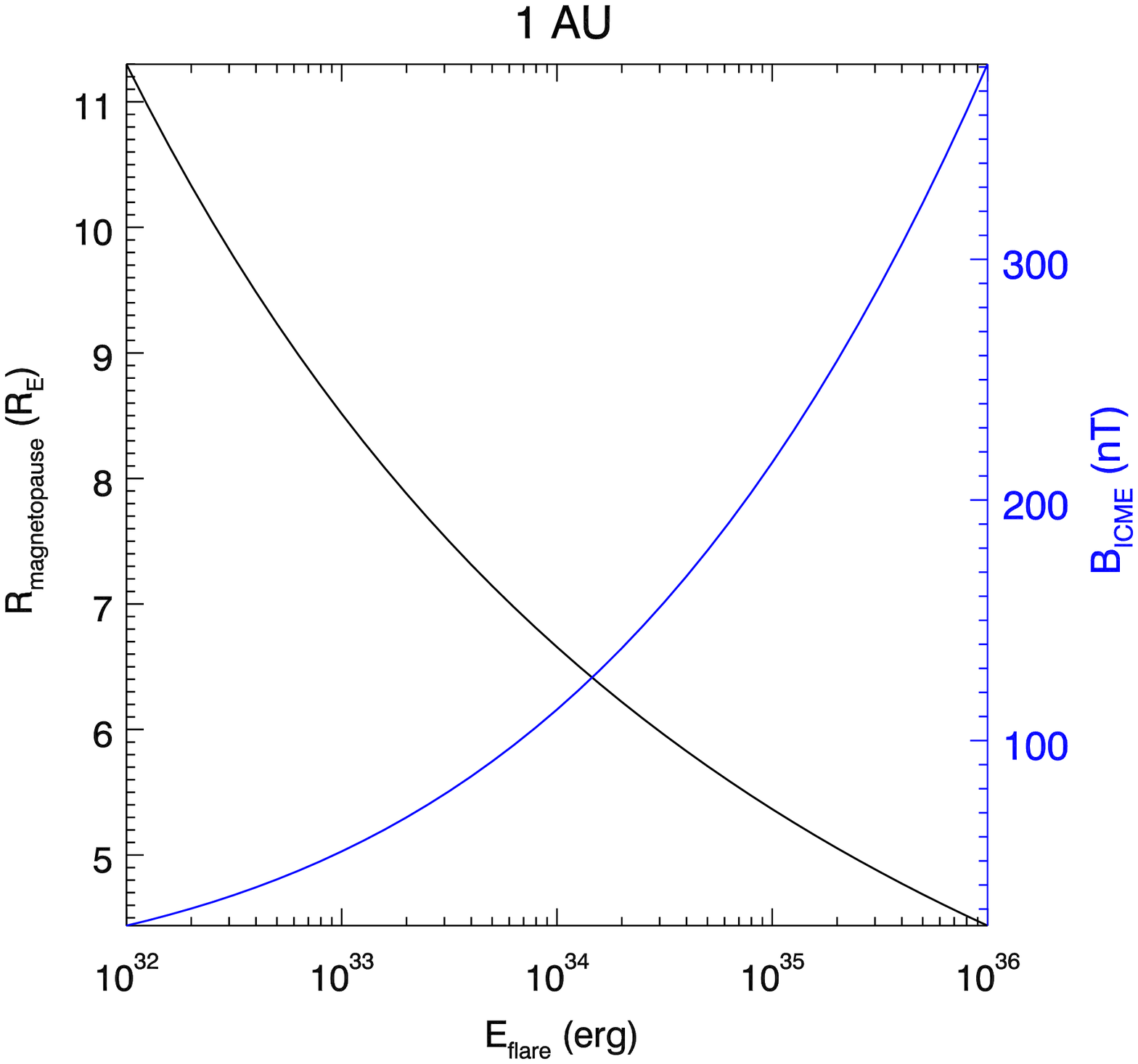}}
\caption{Terrestrial magnetopause radius size (black) and ICME
magnetic field magnitude at 1 AU (blue) as a function of flare
energy.  The index
${\alpha}_{B}$ is equal to -1.9 and the Lundquist model is used.} \label{fig:fig8}
\end{figure}

\begin{figure}
\centerline{\includegraphics[width=0.9\textwidth,clip=]{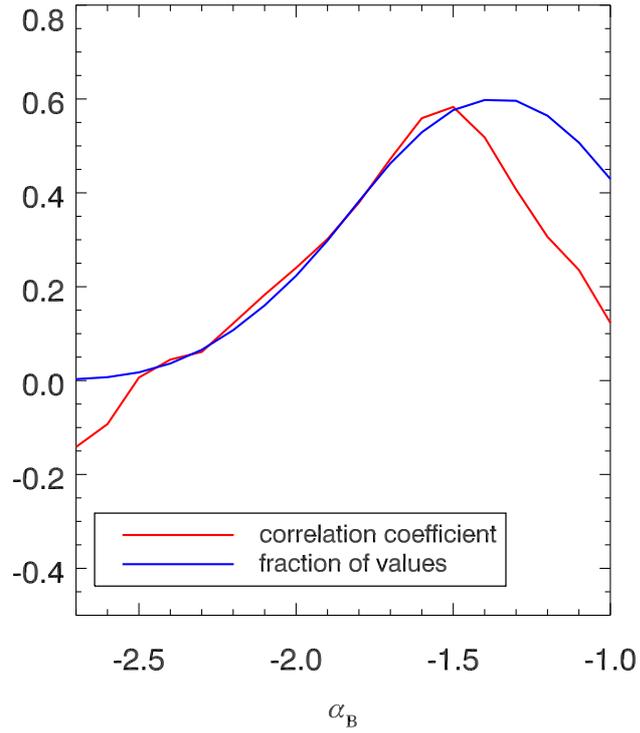}}
\caption{Model-averaged correlation coefficient of the PDF for the predicted $B_\rr{1AU}$ and
observed $B_\rr{MC}$ values at L1 as a function of ${\alpha}_{B}$ (red curve). Also shown is the respective
fraction of ${B_\rr{1 AU}}$ values (blue curve) falling within the
observed $B_\rr{MC}$ range. Calculations shown here are the same as in Figure \ref{fig:fig3} but use a randomly
selected CME $H_\rr{m}$ value in the range $10-40\%$ of the AR $H_\rr{m}$ value (a fixed $100\%$ of the AR $H_\rr{m}$ is used in Figure \ref{fig:fig3}).}
\label{fig:fig9}
\end{figure}

From the discussion above, and  bearing in mind the
corresponding uncertainties, it is reasonable to assume that CMEs, on average, shed $\approx 10-40\,\%$ of the
source region $H_{\rr{m}}$.
To investigate this, we repeat the calculations of Sections 2 and 3, using this time CME $H_\rr{{m}}$
values
randomly selected from 10\,--\,40\,\% of the employed AR $H_\rr{{m}}$ (step 2 in Section 2.1). In Figure \ref{fig:fig9}, we
display  the resulting model-averaged $\rr{{PDF}}_\rr{mod}$\,--\,$\rr{{PDF}}_\rr{obs}$ correlation coefficient
and $\rr{{frac}}_\rr{MC}$  as a function of ${\alpha}_{B}$.
Comparison of the results of Figure \ref{fig:fig9}
and Figure \ref{fig:fig3}, corresponding to the assumption that the entire AR $H_\rr{{m}}$ is attributed
to the associated CME, shows that the curves of Figure \ref{fig:fig9}
are shifted towards larger (\textit{i.e.} flatter) ${\alpha}_{B}$-values by $\approx 0.3$, for both
$\rr{{PDF}}_\rr{mod}$\,--\,$\rr{{PDF}}_\rr{obs}$ and $\rr{{frac}}_\rr{MC}$. Requiring flatter
${\alpha}_{B}$-values for smaller $H_\rr{{m}}$ is reasonably anticipated. This further leads to
smaller near-Sun CME magnetic fields, and therefore the CME magnetic field should evolve less abruptly
with radial distance to match the  MC observations at 1 AU.
Besides this displacement, the overall shapes, peak values and widths of the curves reported
in Figures \ref{fig:fig3} and \ref{fig:fig9} are similar.

Several improvements can be envisioned for both the solar and the stellar
part of our study.
For example, currently our method does not incorporate CME orientation
and rotation in the interplanetary space en route to 1 AU, thereby preventing
the calculation of the vector magnetic field in CMEs. Rather
straightforward-to-implement recipies to incorporate this element exist
\citep[\textit{e.g.}][]{thern2009,wood2010,kay2013,isavnin2014,savani2015}. In addition,
eruption-related helicity changes as found in \citet{tzio2013}, are more appropriate
to use on a case-by-case application of our method, as performed in \citet{pats2016}.
The contribution of magnetic helicity associated with magnetic reconnection
of the erupting flux with its surroundings could be also important \citep[][]{priest2016}.
Better statistics of eruption-related changes are clearly needed.

In the stellar application of our method, a major, largely unconstrained
working hypothesis is that the \citet{tzio2012} relationship between
AR helicity and free-magnetic energy (Equation (\ref{eq:hmefree})),
derived for solar ARs, can be extended to stellar ARs.
This requires futher investigation. For example,
one may derive some rough estimates of stellar AR helicities
by using observations of starspot size and magnetic flux
as derived by Zeeman-Doppler imaging techniques \citep[\textit{e.g.}][]{semel1989,donati2008,vidotto2013}.
Such estimates could be then linked to worst-case flare-energy scenarios
triggered in these stars.
Finally, better constraints, both modeling and observational,
of exoplanet magnetic fields, are required in order to predict more reliably
the influence of stellar CME impacts, and of stellar winds
in general, on exoplanet magnetospheres.

\subsection{Connection to Previous Studies}

Let us now briefly discuss pertinent, complementary efforts to assess
the size of exoplanet magnetospheres around  M-dwarf and Sun-like stars.
\citet{khoda2007} found that the ram-pressure of stellar CMEs could significantly
compress (altitudes $<1000$ km)  the magnetosphere of Earth-like exoplanets, subject to tidal locking,
at distances $< 0.1$ AU around active M stars. By extrapolating
surface magnetic fields of 15 active dM stars, as derived by Zeeman-Doppler imaging, at locations
of exoplanets in their HZ, \citet{vidotto2013} found that Earth-like exoplanets
would require stronger intrinsic magnetic fields
than the terrestrial case, in order to have a magnetosphere with a radius comparable to that of Earth.
Using models and empirical relationships described in \citet[][]{see2014} and \citet{vidotto2014},
\citet{armstrong2016}  found that the extrapolated Kepler 438 to Kepler
438 b stellar wind and magnetic field pressure
leads to a magnetospheric radius for Kepler 438b that is similar to the terrestrial one.
They, however, assumed that Kepler 438b has a magnetic field equal to the terrestrial case,
an assumption which possibly  overestimates the magnetopause radius of Kepler 438b, since
its semi-major axis is small ($\approx 0.16$ AU), so it may experience tidal locking and a decrease in its magnetic field.

Extrapolation of surface magnetic fields combined with stellar
wind models, and extrapolation of assumed stellar CME magnetic fields,
to close-orbit exoplanets around M dwarfs showed that
exoplanet fields of a few tens to hundreds of Gauss, significantly higher than the  terrestrial one,
are required to maintain a magnetopause radius of 2$R_\rr{p}$ \citep{kay2016}.
The bulk of these studies, including ours, in spite of different settings and assumptions, essentially
reach the same basic conclusion, namely, that exoplanets
in the close proximity of their mother stars, could experience significant compression due
to stellar wind and CMEs  that may further hamper their habitability. An example along these lines in our solar system is the recent finding from the Mars Atmosphere and Volatile Evolution (MAVEN) mission team
that ICMEs interacting with the Martian magnetosphere in the early history of the planet
may have played an important role in the planet's atmospheric erosion \citep{jakosky2015}. Here,
of course,
it is not the planet's heliocentric distance -- which is larger than Earth's -- but its weak magnetic field, that has been the major cause for this development. In a related study, \citet{see2014} studied the effect of stellar winds on the sizes of a hypothetical Earth, around  124 Sun-like stars. They found that in most cases, the magnetospheric radius takes values
$>$ 5 terrestrial radii. This is probably a lower limit, because the ram and magnetic pressure of stellar CMEs is not taken into account.

Recent work on the astrosphere of Proxima Centauri and the intrinsic magnetic field
of Proxima b suggest significant compression of their potential magnetospheres,
fully aligned with our results.
3D MHD stellar wind and magnetic field models around
Proxima Centauri, show that the total stellar wind pressure
at Proxima b, could be 2000 times higher than the corresponding one at Earth
\citep{garraffo2016}.
In addition,
planetary evolution models,  scaled to Proxima b,  predict a
 current-day magnetic moment of Proxima b of
$\approx$ 0.32
of the current terrestrial one \citep{zuluaga2016}.
Finally, recent work by \citet{airapet2017} and \citet{dong2017}, using
  MHD simulations, demonstrated the realism of very significant atmospheric losses in Proxima b.

Notice here that tidal locking is achieved for Earth-like
exoplanets around G-M type stars at  distances smaller than  $\approx 0.4-0.7$ AU
\cite[\textit{e.g.} see Figure 2 of][]{griem2005}. Such calculations include only the effect
of gravitational tides. Inclusion of thermal tides linked
to the existence of a planetary atmosphere, internal dissipation effects, eccentric orbits \textit{etc},
could bring exoplanets into asynchronous rotation
\citep[\textit{e.g.}][]{cunha2015,leconte2015}. However, this applies mostly
to exoplanets not very close to their mother stars. \citet{cunha2015}
found for a set of 90 Earth-sized exoplanets
with major semi-axes in the range
0.004-0.54 AU  that they are largely synchronized (see their Table 1).

\section*{Disclosure of Potential
Conflicts of Interest}
The authors declare that they have
no conflicts of interest.

\section*{Acknowledgements}
The authors thank the referee for a useful suggestion to investigate the impact of the uncertainty
of the erupted helicity.
This research has been partly co-financed by the European Union (European
Social Fund -ESF) and Greek national funds through the Operational Program
``Education and Lifelong Learning" of the National Strategic Reference
Framework (NSRF) -Research Funding Program: ``Thales. Investing in knowledge
society through the European Social Fund".
SP acknowledges support from an FP7 Marie Curie
Grant (FP7-PEOPLE-2010-RG/268288).
MKG wishes to acknowledge support  from the EU's Seventh Framework Programme under grant agreement no PIRG07-GA-2010-268245.
The authors acknowledge the  Variability of the Sun and Its Terrestrial Impact
(VarSITI) international program.

\appendix

The Appendix presents a short description of a geometrical CME
model used in the current investigation (A).
Moreover, it contains short descriptions and equations
relating magnetic field magnitude with $H_\rr{m}$ and various geometrical
parameters for  various  theoretical CME models (B\,--\,G). \\

\section{Deducing CME Geometrical Parameters from the GCS Model}
To obtain the geometrical parameters $R$ and $L$ we
adopt the Graduated Cylindrical Shell (GCS) forward fitting model of \citet{thern2009}.
This is a geometrical flux-rope model routinely used
to fit the large-scale appearance
of flux-rope CMEs in multi-viewpoint
observations acquired by coronagraphs onboard the \textit{Solar and Heliospheric
Observatory} (SOHO) and \textit{Solar Terrestrial Relations Observatory} (STEREO) spacecraft.
The GCS user modifies a set of free parameters (CME height, half-angular width $w$,
aspect ratio $k$, tilt angle, central longitude and latitude)
to achieve a best-fit agreement between the model and observations.
A detailed description  can be found in \citet{thern2009}.

In the framework of the GCS model, the CME radius $R$ at a heliocentric distance $r$ is given by the following Equation:
\begin{equation}
R(r)=k r.
\label{eq:cmer}
\end{equation}
To assess the flux-rope length $L$, it is assumed  that the CME front
is  a cylindrical section (see Figure 1 of \citet{dem2009}) with an angular width provided by the geometrical fitting.
One may then write
\begin{equation}
L=2 w r_\rr{mid},
\label{eq:cmelength}
\end{equation}
where $r_\rr{mid}$ is the heliocentric distance half-way through the
model's cross section, along its axis of symmetry. The half-angular width $w$ is given in radians. \\

\section{Cylindrical Linear Force-Free Model}

The Lundquist
flux-rope model \citep{lund1950}
is arguably the most commonly used flux-rope model and corresponds
to a cylindrical force-free solution.

From \citet{dasso2006} we get for a Lundquist flux rope:
\begin{equation}
H_\rr{m}=\frac{4\pi{B_{0}}^{2}L}{\alpha}\int_{0}^{R}{J_{1}}^{2}(\alpha R)\rr{d}r,
\label{eq:hm}
\end{equation}
with $L$  and $R$ the flux-rope length and radius, respectively, $J_{1}$ the Bessel function of the   first
kind, $B_{0}$ the maximum axial field,
and $\alpha$ the force-free parameter.
Making the common assumption of a purely axial  (azimuthal) magnetic field
at the flux-rope axis (edge) we get:
\begin{equation}
\alpha R = 2.405.
\label{eq:alpha}
\end{equation}
\\

\section{Cylindrical Nonlinear Force-Free Model}

This cylindrical nonlinear force-free flux-rope model was proposed
by \cite{goldhoyle1960}.
From  \citet{dasso2006} we have:
\begin{equation}
H_\rr{m}=L(\frac{8\pi[\ln(1+U^{2}/4)]^{2}}{U^{2}})B_{0}^{2}R^{4}{\tau}_{0},
\end{equation}
with $U=2 \tau_{0} R$ and $\tau_{0}=\frac{1}{2}\alpha$. \\

\section{Toroidal Linear Force-Free Model}

This toroidal force-free model was proposed by
\cite{vandas2016}:
\begin{equation}
H_\rr{m}=(2\pi H)\frac{B_{0}^{2}\pi{H}^3}{4{\alpha}_{0}}[8-(1+\frac{1}{{\alpha}_{0}^{2}})\frac{R^{2}}{H^{2}}]J_{1}^{2}(\alpha_{0}),
\end{equation}
with $H=r_\rr{{mid}}$ and $R$ the torus
major and minor axis, respectively, and
$\alpha_{0}=2.405$. R and $r_\rr{{mid}}$ are the same as in Equation (4).\\

\section{Linear Force-Free Model Spheromac}

This linear force-free spheromac model was proposed
by \citet{kataoka2009}:
\begin{equation}
H_\rr{m}=0.045{r_\rr{mid}^{4}}B_{0}^{2},
\end{equation}
and corresponds to a  Sun-centered sphere with radius $r_\rr{{mid}}$ meant
to approximate a spherical magnetic cloud. The value of  $r_\rr{{mid}}$ is the same as in
Equation (4). \\

\section{Cylindrical Constant Current Non Force-Free Model}

This cylindrical constant current
non-force model was
 proposed by \cite{hidalgo2000} and generalized by \cite{nieves2016}.

From \citet{dasso2006} we have  that:
\begin{equation}
H_\rr{m} = \frac{7 \pi}{30} \tau _0 L R^4 B_0^2 ,
\label{eq:b2}
\end{equation}
where  ${\tau}_{0}$ is the  is the twist
per unit length at the flux-rope axis.
The twist parameter ${\tau}_{0}$ can be written as:
\begin{equation}
{\tau}_{0}=\frac{N_\rr{turns}}{L},
\end{equation}
with $N_\rr{turns}$ the total number of field turns along the flux-rope axis.
To estimate ${\tau}_{0}$ we use $L$  as calculated in Section A of Appendix  and assume that $N_{turns}$
is equal to 0.5 and 10,
corresponding to the extreme cases between a weakly and a strongly twisted (multi-turn) flux-rope, respectively.
The number of $N_\rr{{turns}}$  covering  this interval can be deduced from solar imaging and magnetic field observations
(via photospheric magnetic field extrapolations) \citep[\textit{e.g.}][]{vrsn1993,gary2004,guo2013,chintzo2015}
and MC fits at 1 AU \citep[\textit{e.g.}][]{hu2014,wang2016}.
\\

\section{Cylindrical Linear Azimuthal Current Non Force-Free Model}

This cylindrical linear azimuthal current non force-free model was proposed
by \citet{cid2002} and generalized by \cite{nieves2016}. In this model the azimuthal
current increases with distance from the flux-rope axis.

From  \citet{dasso2006} we get:
\begin{equation}
H_\rr{m} = \frac{\pi}{3} \tau _0 L R^4 B_0^2.
\label{eq:b3}
\end{equation}

Like in the  cylindrical constant current non force-free case,
we assume that  $N_\rr{turns}$
is equal to 0.5 and 10.

\bibliographystyle{spr-mp-sola}
\bibliography{refs.bib}

\newpage

\newpage

\newpage

\newpage

\newpage

\newpage

\newpage

\newpage

\newpage

\end{article}
\end{document}